\documentclass[superscriptaddress,twocolumn,pra,aps,showpacs,]{revtex4-1}
\usepackage{epsfig}
\usepackage{bm,hyperref}
\usepackage{amssymb}
\usepackage{mathrsfs}
\usepackage{amsmath}
\usepackage{CJK}

\begin{document}

\title{Collective excitations of a trapped Bose-Einstein condensate in the
presence of weak disorder and a two-dimensional optical lattice}
\author{Ying Hu}
\affiliation{Department of Physics, Centre for Nonlinear Studies, and The Beijing-Hong
Kong-Singapore Joint Centre for Nonlinear and Complex Systems (Hong Kong),
Hong Kong Baptist University, Kowloon Tong, Hong Kong, China}
\author{Zhaoxin Liang}
\email{zhxliang@gmail.com}
\affiliation{Shenyang National Laboratory for Materials Science, Institute of Metal
Research, Chinese Academy of Sciences, Wenhua Road 72, Shenyang 110016, China}
\author{Bambi Hu}
\affiliation{Department of Physics, Centre for Nonlinear Studies, and The Beijing-Hong
Kong-Singapore Joint Centre for Nonlinear and Complex Systems (Hong Kong),
Hong Kong Baptist University, Kowloon Tong, Hong Kong, China}
\affiliation{Department of Physics, University of Houston, Houston, TX 77204-5005, USA}
\date{\today}

\begin{abstract}
We investigate the combined effects of weak disorder and a
two-dimensional (2D) optical lattice on the collective excitations
of a harmonically trapped Bose-Einstein condensate (BEC) at zero
temperature. Accordingly, we generalize the hydrodynamic equations
of superfluid for a weakly interacting Bose gas in a 2D optical
lattice to include the effects of weak disorder. Our analytical
results for the collective frequencies beyond the mean-field
approximation reveal the peculiar role of disorder, interplaying
with the 2D optical lattice and interatomic interaction, on
elementary excitations along the 3D to 1D dimensional crossover. In
particular, consequences of disorder on the phonon propagation and
surface modes are analyzed in detail. The experimental scenario is
also proposed.
\end{abstract}

 \maketitle

\section{introduction}

The measurement of the frequencies of collective excitations has
emerged as a fundamental and precise tool to investigate the quantum
many-body physics of atomic Bose-Einstein condensates (BECs) in
unprecedent detail \cite{Dalfovo,Morschrev,Blochrev,Yukalovrev}. It
not only provides an excellent confirmation of the mean-field
predictions, but also represents a very effective method for probing
the beyond-mean-field effects \cite{Stringari,Pitaevskii,Braaten},
particularly in a weakly interacting Bose gas whose condensate
fraction is close to $100\%$. Theoretically, such a gaseous BEC's
system can be well represented by a single macroscopic wave
function, consequently allowing for clear hydrodynamic formulations
that provide analytic or semi-analytic results in describing dynamic
behavior of a BEC's system.

Along this line, Stringari \cite{Stringari} pioneered in applying
the hydrodynamic equations for superfluid to describe the ground and
excited states of a magnetically trapped dilute interacting Bose gas
at low temperature, yielding predictions in excellent agreement with
the experiment \cite{Dalfovo}. The same scheme was later generalized
by Kr\"amer {\it et al}. \cite{Kramer} to include the significant
effects of optical lattice, following the spectacular experimental
realization of optical lattices \cite{Morschrev,Blochrev}. Recently,
the dramatic effects caused by disorder on the quantum properties of
a BEC have attracted intensive attention \cite{Disorder}. For
example, it has been observed that even a tiny amount of disorder in
the confining fields leads to a fractioning of quasi-1D condensates
in waveguide structures on atom chips \cite{Wang}. One particular
striking effect of disorder arises from Ref. {\cite{Huang}} that
disorder is more active in reducing superfluidity than in depleting
the condensate even at zero temperature ($T=0$). Investigations on
collective modes of a BEC in a random potential using hydrodynamic
scheme have been carried out by Falco {\it {et al.}} in Ref.
\cite{Falco} in lattice-free case. In view of the possibility to
control both disorder and a BEC in an optical lattice almost at will
\cite{Laurent}, an important direction consists in investigating the
collective excitations of a trapped BEC in the combined presence of
disorder and optical lattice.

Another significant influence of disorder on Bose gases is that it
enhances quantum fluctuations and the beyond-mean-field corrections
thereof to the equations of state. Such corrections in the free
space arising from interatomic interaction have been predicted by
Lee, Huang and Yang (LHY) \cite{Yang}. Yet measuring these
corrections remains challenging, as they are usually too small to be
observed. Pitaevskii and Stringari \cite{Pitaevskii} first pointed
out observing these beyond-mean-field effects in the frequency shift
of collective excitations. Later in Ref. \cite{Orso}, Orso {\it{et
al}}. suggested that these frequency shift could be further
magnified by introducing optical lattice to enhance correlations.
Furthermore, Ref. {\cite{Falco}} showed that disorder also caused a
frequency shift in collective modes, with a magnitude comparable to
interatomic interaction. It is therefore expected that the combined
presence of disorder and optical lattice might considerably enhance
the quantum fluctuations, ultimately rendering the effects beyond
mean field more observable in the experiment. On the other hand,
disorder has been shown to cause a departure from Kohn's theorem for
the dipole mode \cite{Falco}, a signature to distinguish the effects
beyond mean field due to disorder from that due to interatomic
interaction. An especially appealing question, therefore, consists
in exploring the beyond-mean-field effects arising from the
interplay between disorder and optical lattice on the collective
excitations of a trapped BEC.

Hence, we are motivated to launch a comprehensive investigation on a
trapped BEC, taking into account of the combined effects of
disorder, optical lattice and interatomic interaction. Against this
background, a 2D optical lattice is especially favorable, because it
induces a characteristic 3D to 1D dimensional crossover in the
quantum fluctuations of a BEC \cite{Orso}, even in the presence of
weak disorder \cite{Hu}. Loading a BEC into a 2D optical lattice,
therefore, provides an effective tool to study a 1D BEC and 3D BEC
through asymptotic analysis \cite{Hu, Orso}.

In this paper, we focus on studying the collective excitations of a
harmonically trapped BEC in the presence of weak disorder and a 2D
optical lattice at $T=0$. Accordingly, we extend the hydrodynamic
equations for superfluid in a 2D optical lattice to include the
effect of weak disorder. Using the developed hydrodynamic equations,
we calculate the collective frequencies beyond the mean-field
approximation. Our analytical results reveal the peculiar role of
disorder, when interplaying with a 2D optical lattice and
interatomic interaction, on the excited states of a trapped BEC
along the 3D to 1D dimensional crossover. In particular, the effects
of disorder on sound velocity and surface modes are discussed in
detail. We also propose possible experimental conditions to realize
our scenario.

The outline of this paper is as follows. In Sec. II, we introduce
the Hamiltonian for a magnetically trapped BEC in the presence of a
2D optical lattice and weak disorder. We first examine the case in
the absence of disorder in Sec. III where we follow the scheme in
\cite{Kramer} for generalizing hydrodynamic equations in Ref.
\cite{Stringari} to include the effects of a 2D optical lattice.
Thereafter in Sec. IV, we account for the presence of random
potential and extend the results in Sec. III to include the effects
of weak disorder. Using these equations, we analyze in Sec. V the
combined effects of disorder and a 2D optical lattice on the
collective excitations of a trapped BEC. In particular, we calculate
the  beyond-mean-field corrections to the collective frequencies.
Detailed discussions on the sound velocity and low-lying modes as
well as quantum behaviors along the dimensional crossover are
presented. Finally in Sec. VI, we summarize our results and propose
the experimental conditions for realizing our scenario.

\section{Hamiltonian for a trapped BEC in the presence of weak disorder and a 2D optical
lattice}

The N-body Hamiltonian describing the Bose system at $T=0$ has the
form \cite{Dalfovo}
\begin{eqnarray}  \label{Ham}
H-\mu N=\!\int
d\mathbf{r}\hat{\Psi}^{\dagger}(\mathbf{r})\Bigg[\!\!&-\!&\!
\frac{\hbar^2\nabla^2}{2m}-\mu\!+\!V_{ext}(\mathbf{r})+V_{ran}(\mathbf{r})
\notag \\
&+&\frac{g}{2}\hat{\Psi}^{\dagger}(\mathbf{r})\hat{\Psi}(\mathbf{r})\Bigg]
\hat{\Psi}(\mathbf{r}),
\end{eqnarray}
where $\hat{\Psi}(\mathbf{r})$ is the field operator for bosons with
mass $m$, $\mu$ is the chemical potential, $\hat{N}=\int d
\mathbf{r}\hat{\Psi}^{\dagger}(\mathbf{r})\hat{\Psi}(\mathbf{r})$ is
the number operator, and $g=4\pi \hbar^2 a/m$ is the coupling
constant with $a$ being the $s$-wave scattering length in the free
space. In Hamiltonian (\ref{Ham}), $V_{ran}(\mathbf{r})$ and
$V_{ext}(\mathbf{r})$ respectively represent the random potential
and the trapping potential.

The external trapping potential $V_{ext}(\mathbf{r})$ in Hamiltonian
({\ref{Ham}}) is generated by a superposition of a 3D harmonic
confinement $V_{ho}(\mathbf{r})$ of magnetic origin and a 2D optical
lattice $V_{opt}(\mathbf{r})$ modulated along the $x-y$ plane
\begin{eqnarray}  \label{Vtr}
V_{ext}(\mathbf{r})
&=&\frac{1}{2}m\left(\omega_x^2x^2+\omega_y^2y^2+\omega_z^2z^2\right)
\nonumber \\
&+&sE_R\left[\sin^2(q_Bx)+\sin^2(q_By)\right],
\end{eqnarray}
where $\omega_x$, $\omega_y$ and $\omega_z$ are the frequencies of
the harmonic trap, $s$ is a dimensionless factor labeled by the
intensity of laser beam and $E_R=\hbar^2q^2_{B}/2m$ is the recoil
energy with $\hbar q_B$ being the Bragg momentum. The lattice period
is fixed by $d=\pi/q_B$.

Disorder $V_{ran}(\mathbf{r})$ in Hamiltonian (\ref{Ham}) is
produced by the random potential associated with quenched impurities
\cite{Huang,Hu,Astra}
\begin{equation}  \label{Ran}
V_{ran}(\mathbf{r})=\sum_{i=1}^{N_{imp}}v\left(|\mathbf{r-r_{i}}|\right),
\end{equation}
with $v(\mathbf{r})$ describing the two-body interaction between
bosons and impurities, $\mathbf{r}_i$ being the randomly distributed
positions of impurities and $N_{imp}$ counting the number of
$\mathbf{r}_i$. To obtain the concrete form of the pair potential
$v(\mathbf{r})$, we need to investigate the scattering problem
between a boson with mass $m$ and a quenched impurity with mass $M$
in the presence of a 2D optical lattice \cite{Wouters}. Here, we
restrict ourself to the conditions of a dilute BEC system in the
presence of a very small concentration of disorder. Thereby, the
potential $v(\mathbf{r})$ can be expressed by an effective
pseudo-potential $v(\mathbf{r })=\tilde{g}_{imp}\delta(\mathbf{r})$
\cite{Huang}. For the sake of notation convenience, one can write
$\tilde{g}_{imp}=2\pi \hbar^2 \tilde{b}/m$ \cite{Huang,Hu,Astra}.
Here, $\tilde{b}$ effectively characterizes the lattice-modified
strength of disorder, while the reduced mass for the boson-impurity
pair coincides with $m$ when the mass of impurity is taken to be
infinite due to its quenched nature \cite{Astra}, i.e.
$\lim_{M\rightarrow\infty}mM/(M+m)=m$.

\section{Macroscopic dynamics of a BEC loaded in a 2D optical lattice}

Let us first consider the absence of disorder ($V_{ran}=0$) in
Hamiltonian (\ref{Ham}). In this section we shall follow the scheme
in Ref. \cite{Kramer} and reproduce corresponding results for
generalizing the hydrodynamic equations of superfluid
\cite{Stringari} to include the effects of a 2D optical lattice at
$T=0$. The validity condition for the presented hydrodynamic
formulation requires relatively large $s$ where the inter-well
barriers are significantly larger than the chemical potential, while
quantum tunneling is still sufficient to ensure full coherence
\cite{Stringari, Kramer}.

Under above conditions, a formulation for the macroscopic dynamics
can be accomplished by considering the long-wavelength limit of
Hamiltonian (\ref{Ham}) and the action functional thereof for the
model system. To this end, we start from the ansatz for the
condensate order parameter within the tight-binding approximation
\cite{Kramer, Orso}
\begin{equation}  \label{Ansatz}
\Phi(\mathbf{r})=\sum_{l_x,l_y}W(x-l_xd)W(y-l_yd)f_{l_x,l_y}(z)e^{iS_{l_x,l_y}\left(z\right)}.
\end{equation}
Here, $S_{l_x,l_y}$ is the phase of the $(l_x,l_y)$ component of the
order parameter, while $W(x-l_xd)$, $W(y-l_yd)$ and $f_{l_x,l_y}(z)$
are real functions. In addition, the $W(x-l_xd)$ and $W(y-l_yd)$ are
further assumed to satisfy the periodic conditions, i.e.
$W(x-l_xd)=W_0(x)$ and $W(y-l_yd)=W_0(y)$ where both $ W_0(x) $ and
$W_0(y)$ are localized at the origin. A transcription to the
continuum limit is achieved by coarse-graining through the
replacement $\sum_{l_x,l_y}\rightarrow (1/d^2)\int dxdy$. One
thereby introduces the smoothed phase $S$ ($S_l(z)\rightarrow
S(\mathbf{r})$) and a coarse-grained density \cite{Kramer}
\begin{equation}\label{Density}
n({\bf{r}})=\frac{1}{d^2}f_{l_x,l_y}^2(z),
\end{equation}
with $x=l_xd$, $y=l_yd$.

The long-wavelength limiting form of the Lagrangian functional can
then be obtained by substituting ansatz ({\ref{Ansatz}}) into
Hamiltonian ({\ref{Ham}}), applying coarse-graining and proceeding
as Ref. \cite{Kramer}:
\begin{eqnarray}
L&=&\langle-i\hbar\int\Phi^{*}\frac{\partial}{\partial t}\Phi d\mathbf{r}+
H-\mu N\rangle  \nonumber \\
&=&\int d\mathbf{r}\Bigg[\hbar \frac{\partial S}{\partial t}-
2t\left[\cos(d\partial_xS)+\cos\left(d\partial_yS\right)\right]  \nonumber \\
&+&\frac{\hbar^2}{2m} \left(\partial_zS\right)^2+ V_{ho}({\bf
r})+\frac{\tilde{g}}{2} n({\bf r})-\mu \Bigg]n({\bf r}),
\label{Lagran}
\end{eqnarray}
where
\begin{equation}
t=-\int
dxW_0(x)\left[-\frac{\hbar^2}{2m}\partial_x^2+V_{opt}(x)\right]W_0(x-d)
\end{equation}
relates to the tunneling rate between adjacent walls with
$V_{opt}(x)=sE_R\sin^2(q_Bx)$ being the external optical lattice
applied along $x$ direction, and $\tilde{g}=4\pi\hbar^2\tilde{a}/m$
is the lattice-renormalized coupling constant with $\tilde{a}=C^2a$
and $C=d\int^{d/2}_{-d/2}W_0^4(x)dx$. In addition, Eq.
({\ref{Lagran}}) is derived based on the Thomas-Fermi (TF)
approximation \cite{Stringari} in the large $N$ limit where one can
neglect the quantum pressure terms originating from the radial term
in the kinetic energy \cite{Kramer}.

With Eq. ({\ref{Lagran}), the equations of motion are derived by
requiring that the action $A=\int_0^t Ldt$ be stationary under
arbitrary variations in the macroscopic density $n({\bf r})$ and the
phase $S(\mathbf{r})$ with fixed endpoints, i.e. $\delta A/\delta
n=0$ and $\delta A/\delta S=0$. The resulting equations are
\cite{Pedri}
\begin{eqnarray}
\frac{\partial n}{\partial t}+\frac{2td}{{\hbar }}\partial _{\perp}\left[n\sin \left(d\partial_\perp S\right)\right]+\frac{\hbar}{m}
\partial _{z}\left[n\partial_z S\right] &=&0,  \label{Hy1} \\
\hbar\frac{\partial S}{\partial t}+\delta
\mu-2t\cos\left(d\partial_{
\perp}S\right)+\frac{\hbar^2}{2m}\left(\partial_z S\right)^2 &=&0,
\label{Hy2}
\end{eqnarray}
where $\delta\mu=V_{ho}+\tilde{g}n-\mu$ is the change of the
chemical potential with respect to its ground state value \cite{Hu}.
In Eqs. ({\ref{Hy1}}) and ({\ref{Hy2}}), two notations have been
introduced for simplicity: $\partial _{\perp}\left [n\sin
\left(d\partial_\perp S\right)\right ]\equiv \partial _{x}\left
[n\sin \left(d\partial_x S\right)\right ]+\partial _{y}\left [n\sin
(d\partial_y S)\right ]$ and $\cos\left(d\partial_{\perp}S\right)
\equiv
\cos\left(d\partial_{x}S\right)+\cos\left(d\partial_{y}S\right)$.
Accordingly, one can define the superfluid velocity $\mathbf{v}$ in
a 2D optical lattice as \cite{Kramer}
\begin{eqnarray}
\upsilon _{x(y)} &=&\frac{\hbar }{m^{\ast }}\partial _{x(y)}S,  \label{vx} \\
\upsilon _{z} &=&\frac{\hbar }{m}\partial _{z}S,\label{vz}
\end{eqnarray}
where the effective mass $ m^*$ is introduced by
\begin{equation}
\frac{m}{m^*}=\frac{2mtd^2}{\hbar^2},
\end{equation}
which accounts for the increased inertia of the system along the
direction of optical lattice \cite{Kramermass,Liang}. The ground
state density within the TF approximation is obtained by putting
$\delta \mu=0$ that yields \cite{Stringari, Kramer}:
\begin{equation}\label{NTFL}
n_{TF}({\bf r})=\frac{1}{\tilde{g}}\left[ \mu -\frac{1}{2}m\left(
\omega _{x}^{2}x^{2}+\omega _{y}^{2}y^{2}+\omega
_{z}^{2}z^{2}\right) \right].
\end{equation}
Eq. ({\ref{NTFL}}) shows a fattened cloud in the presence of a
lattice due to increased repulsive interaction characterized by
$\tilde{g}$.

In the limit of small oscillations, Eqs. (\ref{Hy1}) and (\ref{Hy2})
can be linearized by substitutions of Eqs. (\ref{vx}) and (\ref{vz})
and decomposing $n(\mathbf{r})=n_{TF}(\mathbf{r})+\delta
n(\mathbf{r})$ and $\mathbf{v}(\mathbf{r})=\delta \mathbf{v}
(\mathbf{r})$. The resulting equations read,
\begin{eqnarray}
&&\frac{\partial \delta n}{\partial t}+\nabla \cdot \lbrack n_{TF}\delta
\mathbf{v}]=0,  \label{Hy3} \\
&&\hbar \frac{\partial \delta S}{\partial t}+\delta \mu =0.  \label{Hy4}
\end{eqnarray}
With Eqs. ({\ref{vx}}) and ({\ref{vz}}), one can combine Eqs.
(\ref{Hy3}) and ({\ref{Hy4}}) and consequently obtain the equation
for the propagation of a density disturbance in the model system at
$T=0$,
\begin{equation}
m\frac{\partial ^{2}\delta n}{\partial t^{2}}-\widetilde{\nabla
}\cdot\left[ n_{TF} \widetilde{\nabla }\delta \mu \right] =0,
\label{neff}
\end{equation}
where we have introduced the notion \cite{Orso}
\begin{eqnarray}
\widetilde{\nabla }\equiv \left(\widetilde{\nabla }_{\perp},\nabla_z\right)
=\left(\sqrt{\frac{m}{m^{\ast }}}\frac{\partial }{
\partial x},\sqrt{\frac{m}{m^{\ast }}}\frac{\partial }{\partial y},\frac{
\partial }{\partial z}\right).  \label{nablaeff}
\end{eqnarray}

Eq. ({\ref{neff}}) bears formal resemblance with the corresponding
equation in Ref. \cite{Stringari} for small-amplitude density
fluctuation in the free space. Hence by assuming that the effect of
optical lattice is captured by the effective mass $m^{*}$ and the
renormalized interaction $\tilde{g}$, one can study the quantum
behavior of a BEC as if the space is homogeneous in the absence of
harmonic trap, despite the application of periodic potential. In
particular, in the limit $8t\gg\mu$ where the system retains a 3D
behavior, Ref. \cite{Fort} has experimentally verified the validity
of the above mass renormalization theory. In the opposite 1D regime,
on the other hand, the use of hydrodynamic equations is justified by
the superfluidity of the quasi-1D tubes \cite{Moritz} and by the
avoidance of the Mott-insulator phase.

\section{Macroscopic dynamics of a BEC in the presence of weak disorder and a 2D optical
lattice}

In this section, we take into account of $V_{ran}({\bf r})\neq 0$ in
Hamiltonian (\ref{Ham}) and investigate the effects of weak
disorder, combined with a 2D optical lattice, on elementary
excitations of a trapped BEC at $T=0$. To this purpose, we
generalize corresponding hydrodynamic scheme in Sec. III to include
the effects of weak external randomness.

The disorder in this paper is associated with quenched impurities
which generate a random potential $V_{ran}({\bf r})$ given by Eq.
(\ref{Ran}). For simplicity, we further assume that the randomness
is uniformly distributed with density $n_{imp}=N_{imp}/V$ and
Gaussian correlated \cite{Astra}. Hence, the two basic statistical
properties of disorder, in the presence of a 2D optical lattice, are
the average value $\langle V_0\rangle=\tilde{g}_{imp}n_{imp}$ and
the correlation function
\begin{equation}\label{Vkvk}
\langle V_{ \mathbf{k}}
V_{-\mathbf{k}}\rangle=\frac{1}{V}\tilde{g}^2_{imp}n_{imp},
\end{equation}
where $\tilde{g}_{imp}=2\pi\hbar^2\tilde{b}/m$ is the
lattice-renormalized boson-impurity interaction, $n_{imp}$ is the
impurity concentration, and $ V_{\mathbf{k}}=\left(1/V\right)\int
e^{i\mathbf{k}\cdot \mathbf{r}}V_{ran}( \mathbf{r})d\mathbf{r}$ is
the Fourier transform of $V_{ran}(\mathbf{r})$. Here, the notation
$\langle ... \rangle$ stands for the ensemble average over all
possible disorder configurations \cite{Yukalovd}. The effect of
disorder in a lattice can be characterized by two important
parameters: the concentration of disorder $\kappa=n_{imp}/n$ and the
ratio of effective interaction strength
\begin{equation}\label{R}
\tilde{R}=\frac{n_{imp}}{n} \frac{\tilde{b}^2}{\tilde{a}^2}.
\end{equation}

The validity conditions for following hydrodynamic formulation in a
random potential require the correlation length of disorder to be
much smaller than the healing length of superfluid \cite{Falco}, in
addition to those required in Sec. III.

\subsection{Hydrodynamic equations}

We start from Hamiltonian ({\ref{Ham}}) with $V_{ran}({\bf r})$
given by Eq. ({\ref{Ran}) within the white noise approximation
\cite{Astra}, and follow similar procedures in Sec. III. The
resulting hydrodynamic equation for the macroscopic phase $S$ reads
\begin{equation}  \label{Hytwo1}
\hbar\frac{\partial S}{\partial t}+\delta
\mu-2t\left[\cos\left(d\partial_{ x}S\right)+\cos\left(d\partial_{
y}S\right)\right]+\frac{\hbar^2}{2m}\left(\partial_z S\right)^2 =0,
\end{equation}
where $\delta \mu=V_{ho}+\tilde{g}n-\mu$ is the change of chemical
potential with respect to its ground state value. Compared to Eq.
(\ref{Hy2}) in the absence of disorder ($V_{ran}=0$), Eq.
(\ref{Hytwo1}) takes similar form, except that the associated
chemical potential $\mu$ here includes the dramatic effects of
disorder on the system's ground state \cite{Huang,Giorgini,Hu}. The
presence of disorder, therefore, only affects the dynamics of phase
fluctuation indirectly through the equation of state. We will
determine $\mu$ in the following section. Once Eq. ({\ref{Hytwo1}})
is solved for $S$, the superfluid velocity $\mathbf{v}$ can be
obtained from Eqs. ({\ref{vx}}) and (\ref{vz}). In the limit of
small-amplitude phase fluctuation $S=\delta S$, Eq. ({\ref{Hytwo1}})
can be linearized as
\begin{equation}\label{dv_1}
\hbar\frac{\partial \delta S}{\partial t}+\delta \mu=0.
\end{equation}
Here, the hydrodynamical variables are assumed to be ensemble
averages over all possible realizations of external disorder. By
this assumption, the validity of Eqs. (\ref{Hytwo1}) and
(\ref{dv_1}) should be restricted to the self-averaging regime where
the wavelength of the hydrodynamic modes is much larger than the
correlation length of disorder \cite{Falco}.

For the equation of total density $n({\bf r})$, the corresponding
conservation law is unaffected by the presence of weak disorder.
However, the random potential gives rise to a normal component in
the fluid \cite{Huang,Hu,Falco,Giorgini} which is pinned by quenched
impurities and is neither a fluid nor dynamic. Thus, when a small
perturbation $\delta {\bf v}$ is applied, only superfluid component
responds to the probe and contributes to the induced current ${\bf
j}=n_s\delta{\bf v}$ with $n_s$ being the superfluid density
\cite{Giorgini,Yukalovd}, whereas the normal component remains
stationary. Consistent with this picture, a density disturbance
$\delta n({\bf r})$ from the equilibrium density profile is related
to the perturbation $\delta\mathbf{v}$ through the equation of
continuity
\begin{equation}
\frac{\partial \delta n}{\partial t}+\nabla \cdot \left[ n_{s}\delta
\bf{v}\right] =0. \label{dn_1} \\
\end{equation}
Here, $\delta{\bf v}$ is determined from $\delta S$ through Eqs.
({\ref{vx}}) and ({\ref{vz}}). Hence Eq. ({\ref{dn_1}}) can be
combined with Eq. ({\ref{dv_1}}) to yield the hydrodynamic equation
\begin{equation}  \label{dn_3}
m\frac{\partial^2 \delta n}{\partial t^2}+\tilde{\nabla} \cdot
\left[ n_{s} \tilde{\nabla}\delta \mu\right] =0,
\end{equation}
where the notation $\tilde{\nabla}$ is given by Eq.
({\ref{nablaeff}}).

Eq. ({\ref{dn_3}}) is one of the main results of this paper. This
equation is characterized by the disorder-depleted superfluid
density $n_s$, the renormalized interaction $\tilde{g}$ and an
affective mass $m^*$ accounting for the presence of optical lattice.
For vanishing disorder, the whole system participates superflow and
Eq. (\ref{dn_3}) becomes the corresponding equation for superfluid
in Refs. \cite{Orso,Kramer}; whereas for vanishing optical lattice,
our result recovers the corresponding one in Ref. \cite{Falco};
while for vanishing optical lattice as well as disorder, Eq.
(\ref{dn_3}) passes to that in Ref. \cite{Stringari} for superfluid
in harmonic traps. We emphasize that Eq. ({\ref{dn_3}}) has no
similarity with the corresponding phenomenological equations in
Landau's two-fluid theory \cite{Landau}. The disorder induced normal
component is not dynamic. It's effect on the dynamics of superfluid
is indirectly exhibited through the superfluid density $n_s=n-n_n$
and the equation of state.

\subsection{Equations of state beyond the mean-field approximation}

Eq. ({\ref{dn_3}}) needs to be supplemented with prescriptions for
the chemical potential $\mu(\mathbf{r})$, the ground state density
$n(\mathbf{r})$ and the superfluid density $n_s(\mathbf{r})$. In
what follows, we derive expressions for these quantities based on
the local density approximation (LDA) \cite{Stringari}.

\subsubsection{Chemical potential}

For the density profile that varies on a macroscopic scale, one can
invoke the LDA for the chemical potential
\begin{equation}
\mu (\mathbf{r})=\mu _{l}[n(\mathbf{r})]+V_{ho}(\mathbf{r}),  \label{mul}
\end{equation}
where $\mu _{l}$ refers to the chemical potential calculated for a
BEC's system with Hamiltonian (\ref{Ham}) in the absence of harmonic
trap ($V_{ho}=0$).

The $\mu_l$ can be determined from $\mu_l =\partial E_{g}/\partial
N$, with $E_{g}$ representing the ground state energy of the model
system without harmonic trap. To calculate $E_{g}$ microscopically,
we adopt Bogoliubov's theory \cite{Hu} and expand the field
operators in Hamiltonian (\ref{Ham}) in the form $\hat{
\Psi}(\mathbf{x})=\sum_{\mathbf{k}}\hat{a}_{\mathbf{k}}e^{-ik_{z}z}\phi
_{{ k_{x}}}\left( x\right) \phi _{{k_{y}}}\left( y\right)$, where
$\phi _{k_{x}}(x)\phi _{k_{y}}(y)$ corresponds to the lowest Bloch
band of the BEC system under consideration. In the tight-binding
approximation, one can write
$\phi_{k_{x}}(x)=\sum_{l}e^{ilk_x}w(x-ld)$ with $w(x)=\exp
[-x^{2}/2\sigma ^{2}]/\pi ^{1/4}\sigma ^{1/2}$ and $d/\sigma \simeq
\pi s^{1/4}\exp (-1/4 \sqrt{s})$ \cite{Kramer,Hu}.  The dominate
presence of condensate is taken into account through the
approximation $\hat{a}_{\mathbf{k}}\approx \sqrt{nV}\delta
_{\mathbf{k}0}+\delta \hat{a}_{\mathbf{k}}$. Retaining only the
quadratic terms in the excitations $\delta \hat{a}_{ \mathbf{k}}$
and $\delta \hat{a}_{\mathbf{k}}^{\dag }$ from the condensate, the
truncated Hamiltonian (\ref{Ham}) can be ultimately diagonalized by
the Bogoliubov transformation \cite{Hu}:
$\hat{a}_{\mathbf{k}}=u_{\mathbf{k}}\hat{c}_{ \mathbf{k}}-\upsilon
_{\mathbf{k}}\hat{c}_{-\mathbf{k}}^{\dag }-\sqrt{N
}V_{k}\left(u_{\mathbf{k}}-\upsilon
_{\mathbf{k}}\right)^{2}/E_{\mathbf{k}}$ and $
\hat{a}_{\mathbf{k}}^{\dagger
}=u_{\mathbf{k}}\hat{c}_{\mathbf{k}}^{\dagger }-\upsilon
_{\mathbf{k}}\hat{c}_{-\mathbf{k}}-\sqrt{N}V_{-k}\left(u_{\mathbf{k}
}-\upsilon _{\mathbf{k}}\right)^{2}/E_{\mathbf{k}}$ with $\upsilon
_{ \mathbf{k}}^{2}=u_{\mathbf{k}}^{2}-1=\left[\left(\varepsilon
_{k}^{0}+\tilde{g}n\right)/E_{\mathbf{k}}-1\right]/2$ and
$E_{\mathbf{k}}=\sqrt{ \varepsilon _{k}^{0}\left(\varepsilon
_{k}^{0}+2\tilde{g}n\right)}$. Here, $ V_{\mathbf{k}}$ is the
Fourier transform of $V_{ran}(\mathbf{r})$ and $\epsilon_{{\bf
k}}^{0}=\hbar^2k^2_z/2m+2t[2-\cos(k_xd)-\cos(k_yd)]$ is the energy
dispersion of a noninteracting Bose gas in the presence of a 2D
optical lattice. Consequently, one finds the ground state energy
$E_{g}=\tilde{g}n^{2}V/2 -\sum_{k\neq 0}(\varepsilon _{\bf
k}^{0}+\tilde{g}n-E_{\bf
k})/2+N\left[\tilde{g}_{imp}n_{imp}-\left(\tilde{g}_{imp}^{2}n_{imp}/V\right)
\sum_{\bf k\neq 0}\varepsilon _{\bf k}^{0}/E_{\bf k}^{2}\right]$.

The resulting $\mu_l$ in Eq. ({\ref{mul}}), in a convenient form, is
given by
\begin{equation}
\mu _{l}(n)=\tilde{g}n\left[ 1+k_{int}(n)+k_{dis}(n)\right],
\label{mul_1}
\end{equation}
which includes the first order correction to the result
$\mu=\tilde{g}n$ in the mean-field approximation. Precisely,
interatomic interaction and weak disorder respectively give rise to
the beyond-mean-field corrections
\begin{eqnarray}
k_{int}(n)&=&\frac{1}{4\pi\hbar}\frac{\sqrt{2m\tilde{g}n}}{nd^2}\left[x
\frac{d f(x)}{dx}-\frac{3}{2}f\left(x\right)\right]
\label{kin}
\end{eqnarray}
and
\begin{eqnarray}
k_{dis}(n)&=&\kappa\frac{\tilde{b}}{\tilde{a}}\!+\!\frac{\tilde{R}}{4\pi\hbar}\frac{\sqrt{m\tilde{g}n}}{nd^2}\left[x
\frac{d Q(x)}{dx}\!-\!\frac{3}{2}Q\left(x\right)\right], \label{kd}
\end{eqnarray}
with $x=2t/\tilde{g}n$ and $\kappa=n_{imp}/n$. Here, the $f(x)$ and
$Q(x)$ as functions of variable $x$ are respectively defined as
\cite{Hu}
\begin{equation}
f(x)=\frac{\pi }{2\sqrt{x}}\int_{-\pi }^{\pi }\frac{d^{2}\mathbf{\mathbf{k}}
}{(2\pi )^{2}}\frac{_{2}F_{1}\left(\frac{1}{2},\frac{3}{2},3,-\frac{2}{x{\gamma (
\mathbf{k})}}\right)}{\sqrt{\gamma (\mathbf{k})}},  \label{f}
\end{equation}
and
\begin{equation}
Q(x)=\frac{\pi }{2}\int_{-\pi }^{\pi }\frac{dk_{x}}{2\pi }\frac{_{2}F_{1}\left(
\frac{1}{2},\frac{1}{2},1,\frac{x^{2}}{\left( 1+x+x\sin ^{2}(k_{x}/2)\right)
^{2}}\right)}{\sqrt{x\sin ^{2}\left( k_{x}/2\right) +1}}.  \label{Q}
\end{equation}
In Eq. (\ref{f}), $\gamma (\mathbf{k})=2-\cos (k_{x})-\cos (k_{y})$.
The function $ _{2}F_{1}(a,b,c,d)$ in Eqs. (\ref{f}) and (\ref{Q})
is the hypergeometric function \cite{Zwillinger} and the integration
over the transverse quasimomenta is restricted to the first
Brillouin zone, i.e. $|k_{x}|,|k_{y}|\leq \pi$ in Eq. (\ref{f}) and
$|k_{x}|\leq \pi $ in Eq. ( \ref{Q}).

The parameter $x=2t/\tilde{g}n$ controls the dimensional crossover
for a uniform gas ($V_{tr}=0$). In the limit $x\rightarrow 0$,
corresponding to $8t\ll \mu_l$, the system undergoes a dimensional
crossover to a 1D regime where the $f(x)$ saturates to the value
$4\sqrt{2}/3$. In this limit, we can neglect the Bloch dispersion
and Eq. (\ref{mul_1}) approaches asymptotically to the chemical
potential of a 1D Bose gas in the presence of weak disorder
\begin{eqnarray}
\mu _{l}(n)&=&\tilde{g}n\Bigg\{1-\frac{1}{\pi \hbar}\frac{\sqrt{m\tilde{g}n}}{nd^2}\nonumber\\
&+&\kappa\frac{\tilde{b}}{\tilde{a}}
+\frac{\tilde{R}}{4\pi\hbar}\frac{\sqrt{m\tilde{g}n}}{nd^2}\left[x
\frac{d Q(x)}{dx}-\frac{3}{2}Q\left(x\right)\right]\Bigg\}.\label{E1D}
\end{eqnarray}
Eq. ({\ref{E1D}}) effectively generalizes the Lieb-Linger solution
of the 1D model expanded in the weak coupling regime to include the
effects of weak disorder \cite{Hu}.

In the opposite limit $x \gg 1$, corresponding to $8t\gg \mu_l$, the
system retains an anisotropic 3D behavior. In this regime, the
functions (\ref{f}) and (\ref{Q}) respectively reach the asymptotic
law $f(x)\simeq 1.43/\sqrt{x}-16\sqrt{2}/15\pi x$ and $Q(x)\simeq
-1/\sqrt{\pi}x$. Eq. (\ref{mul_1}) thus takes the asymptotic form
\begin{eqnarray}
\mu_l(n)=\tilde{g}n\Bigg[\left(1+\kappa
\frac{\tilde{b}}{\tilde{a}}+\frac{\tilde a}{\tilde
a_{cr}}\right)&+&\frac{32}{3\sqrt{\pi}}\frac{m^{*}}{m}\sqrt{n\tilde{a}^3}\nonumber
\\
&+&5\tilde{R}\frac{m^{*}}{m}\sqrt{n\tilde{a}^3}\Bigg].\label{E3D}
\end{eqnarray}
In Eq. ({\ref{E3D}}), the $(\kappa\tilde{b})/\tilde{a}$ and
$\tilde{a}/\tilde a_{cr}=-(0.24\tilde{a}/d)\sqrt{m^{*}/m}$
\cite{Orso} provide a further renormalization of the scattering
length due to the random potential and optical lattice, i.e.
$\tilde{g}_{eff}=\tilde{g}\left(1+\kappa \tilde{b}/\tilde{a}+\tilde
a/\tilde a_{cr}\right)$; whereas, the remaining two terms
proportional to the renormalized gas parameter generalize the LHY
correction \cite{Yang} to include effects of optical lattice and
disorder. Eq. (\ref{E3D}) therefore shows that the effects beyond
mean field are much amplified via the introduction of periodic
potential and external randomness, due to the increased inertia of
mass $m^{*}$ along the direction of lattice and the interaction of
bosons with impurities characterized by $\tilde{R}$. For vanishing
disorder $\kappa=0$, Eq. (\ref{E3D}) is reduced to the corresponding
result in Ref. \cite{Orso}; whereas for vanishing optical lattice
$s=0$, our result recovers exactly the corresponding one in Ref.
\cite{Astra}.

\subsubsection{Ground state density profile}
Using Eq. ({\ref{mul_1}}), one obtains by iteration the LDA ground
state density profile $n({\bf r})$ \cite{Stringari}
\begin{equation}  \label{nbey}
n({\bf r})=n_{TF}-n_{TF}k_{int}(n_{TF})-n_{TF}k_{dis}(n_{TF}),
\end{equation}
where
\begin{equation}\label{Ntf_1}
n_{TF}(\mathbf{r})=n(0)-\frac{m}{2\tilde{g}}(\omega_x^2x^2+\omega_y^2y^2+
\omega_z^2z^2)
\end{equation}
is the mean-field value of density within the TF approximation. In
Eq. ({\ref{Ntf_1}}), $n(0)$ is the lattice-renormalized TF density
evaluated at the center of harmonic trap determined from
\begin{equation}\label{N0}
n(0)=\frac{m}{2\tilde{g}}(\omega_x^2X^2+\omega_y^2Y^2+
\omega_z^2Z^2),
\end{equation}
where $X, Y,Z$ denote the size of the system. The second and third
terms in Eq. ({\ref{nbey}}) show that the quantum fluctuations due
to interatomic interaction and disorder reduce the LDA total
density, thereby fattening the cloud.

In the asymptotic 3D regime, $k_{int}$ in Eq. (\ref{kin}) and
$k_{dis}$ in Eq. (\ref{kd}) respectively approach
\begin{equation}\label{Kint3D}
k_{int}({\bf
r})=\frac{\tilde{a}}{\tilde{a}_{cr}}+\frac{32}{3\sqrt{\pi}}\frac{m^*}{m}(n\tilde{a})^{\frac{3}{2}},
\end{equation}
 and
\begin{equation}\label{Kdis3D}
k_{dis}({\bf r})=\kappa\frac{\tilde {b}}{\tilde
{a}}+5\tilde{R}\frac{m^*}{m}(n\tilde{a})^{\frac{3}{2}}.
\end{equation}
It thus follows from Eq. ({\ref{nbey}}) that the total density in
the 3D limit takes the form
\begin{equation}\label{nbey1}
\!n({\bf r})=n_{TF}\left(\!1-\!\kappa
\frac{\tilde{b}}{\tilde{a}}\!-\frac{\tilde a}{\tilde
a_{cr}}\right)-\left(\frac{32}{3\sqrt{\pi}}+5\tilde{R}\!\right)\frac{m^{*}}{m}\left(\tilde{a}n_{TF}\right)^{\frac{3}{2}}.
\end{equation}
Eq. ({\ref{nbey1}}) shows that the effects beyond mean field due to
both optical lattice and disorder have two consequences on the total
density: first of all the TF density is further renormalized arising
from the modified effective scattering length $n_{TF}\left(1-\kappa
\tilde{b}/\tilde{a}-\tilde a/\tilde a_{cr}\right)\simeq
\tilde{g}n_{TF}/\tilde{g}_{eff}=n_{eff}$; while the second point
concerns the generalized LHY correction in the presence of disorder
and optical lattice as exhibited by the last term in Eq.
({\ref{nbey1}}). For vanishing disorder and optical lattice, Eq.
({\ref{nbey1}}) is reduced to the corresponding one in Ref.
\cite{Pitaevskii}.

It's important to mention here that the density $n({\bf r})$
entering the hydrodynamic equations is the total density and shall
not be confused with the condensate density $n_c({\bf r})$
\cite{Pitaevskii}, particularly in cases where the effects beyond
mean field are taken into account. In the asymptotic 3D regime, by
extending the result of quantum depletion in \cite{Hu} for an
effectively uniform gas to that with harmonic trap using LDA, and
together with Eq. ({\ref{nbey1}) for the total density, one obtains
the condensate density of the model system
\begin{eqnarray}\label{Nc}
n_c({\bf r})&=&n_{TF}\left(1-\kappa
\frac{\tilde{b}}{\tilde{a}}-\frac{\tilde a}{\tilde
a_{cr}}\right)\nonumber\\
&-&\left[\frac{40}{3\sqrt{\pi}}+(5+\frac{\sqrt{\pi}}{2})\tilde{R}\right]\frac{m^{*}}{m}\left(\tilde{a}n_{TF}\right)^{\frac{3}{2}}.
\end{eqnarray}
Comparison of Eqs. ({\ref{nbey1}}) and ({\ref{Nc}}) shows that the
effects beyond mean field reduce more condensate density than the
total density. The $n({\bf r})$ and $n_c({\bf r})$ only equal in the
limit where quantum depletions are negligible. In particular, Eq.
({\ref{Nc}}) demonstrates that the disorder induced quantum
depletion
\begin{equation}\label{NCd}
n_R({\bf
r})=\kappa\frac{\tilde{b}}{\tilde{a}}n_{TF}+(5+\frac{\sqrt{\pi}}{2})\tilde{R}\frac{m^{*}}{m}\left(\tilde{a}n_{TF}\right)^{\frac{3}{2}}
\end{equation}
is considerably amplified at each point of the space, compared to
the corresponding result in Ref. \cite{Huang}, when
beyond-mean-field effects are included.

\subsubsection{Superfluid density profile}

The superfluid density is determined from $n_s({\bf r})=n({\bf
r})-n_n({\bf r})$. Here, $n({\bf r})$ is the total density
determined from Eq. (\ref{nbey}), while $n_n({\bf r})$ is the normal
fluid density. For a uniform gas in the absence of harmonic trap,
the normal density $n_n$ can be determined from the long-wavelength
limit of the static transverse current-current response function
$\chi_{T}(\mathbf{ q})$ at $T=0$ \cite{Huang,Forster,Liquid,Baym}
\begin{equation}  \label{Normal}
n_n =\lim_{q\rightarrow 0}\chi_{T}(\mathbf{q}),
\end{equation}
where $\chi_{T}(\mathbf{q})$ can be derived within Bogliubov's
theoretical framework \cite{Huang,Falco}. The result for $n_n$ of a
BEC in the presence of weak disorder and a 2D optical lattice has
been derived in our previous work \cite{Hu} within the context of
renormalized mass theory. Thus by extending the corresponding result
in Ref. \cite{Hu} to that with harmonic trap using LDA, we obtain
\begin{equation}\label{NFluid}
n_n({\bf r})=\frac{\tilde{R}}{4\hbar
d^2}\sqrt{2m\tilde{g}n_{TF}}I(\frac{2t}{\tilde{g} n_{TF}}).
\end{equation}
Here, the $n_{TF}$ is the position-dependent TF density given by Eq.
({\ref{Ntf_1}}) and the function $I(x)$ with $x=2t/\tilde{g}n_{TF}$
is defined as
\begin{equation}  \label{I}
I(x)\!=\!\int_{-\pi }^{\pi
}\frac{d^2\mathbf{k}}{(2\pi)^2}\frac{1}{\sqrt{x\gamma(\mathbf{k})
+2}\left[ \sqrt{x}+\sqrt{x\gamma(\mathbf{k})+2}\right] ^{2}},
\end{equation}
where $\gamma (\mathbf{k})=2-\cos (k_{x})-\cos (k_{y})$. Here, the
parameter $x=2t/\tilde{g}n_{TF}$ controls the dimensional crossover
for nonuniform gases. In the limit of 3D regime corresponding to
$x\gg 1$, $I(x)$ asymptotically approaches the function
$\sqrt{2}/(6\pi x)$. Accordingly, the LDA normal fluid density
(\ref{NFluid}) takes the asymptotic form
\begin{equation}\label{Normal3D}
n _{n}({\bf r})=\frac{2\sqrt{\pi}}{3} \tilde{R} \frac{m^{\ast
}}{m}n_{TF}\left(n_{TF}\tilde{a}^3\right) ^{\frac{1}{2}}.
 \end{equation}
However, with the disorder induced quantum depletion given by Eq.
({\ref{NCd}}), the ratio $n_n/n_R$ at each point of the space
departures from the factor $4/3$ pointed out in Ref. \cite{Huang},
as a result of the effects beyond mean field.

\section{Collective excitations}

Having at hand all the quantities needed to solve Eq.
({\ref{dn_3}}), we can now proceed to investigate the low-energy
collective excitations of the BEC's system under consideration. For
this purpose, it's more advantageous to write $n_s=n-n_n$ and
$\delta \mu=(\partial\delta \mu_l/\partial n_{TF})\delta n$ in Eq.
({\ref{dn_3}}) with $\mu_l$ given by Eq. (\ref{mul_1}). After some
elaborated algebra, we ultimately obtain
\begin{eqnarray}
m\omega^2\delta n&+&\tilde{\nabla}\cdot \left[
n_{TF}\tilde{\nabla}\tilde{g}\delta
n\right]=-\tilde{\nabla}^{2}\left(
\tilde{ g}n_{TF}^{2}\frac{\partial k_{int}}{\partial n_{TF}}\delta n\right)\nonumber \\
&-&\tilde{\nabla}^{2}\left( \tilde{ g}n_{TF}^{2}\frac{\partial
k_{dis}}{\partial n_{TF}}\delta n\right)+\tilde{\nabla}\cdot \left[
n_{n}\tilde{\nabla}\tilde{g}\delta n\right]  .  \label{bmean2}
\end{eqnarray}
Eq. ({\ref{bmean2}}) provides the appropriate generation of the
zeroth order hydrodynamic equation in a 2D optical lattice
\cite{Kramer}
\begin{equation}  \label{mean}
m\omega^2\delta n+\tilde{\nabla}\cdot \left[
n_{TF}\tilde{\nabla}\tilde{g}\delta n\right] =0,
\end{equation}
where the frequency $\omega$ can be solved for low-energy collective
excitations within the mean-field approximation \cite{Stringari,
Kramer}. After solutions to Eq. ({\ref{mean}}) are found, one can
then solve Eq. (\ref{bmean2}) by treating all the terms on the right
side as perturbation. Two types of such beyond-mean-field
corrections can be identified: the first is associated with
increased bulk compressibility $\chi^{-1}=\partial \mu/\partial n$
due to the enhanced quantum fluctuations caused by the combined
effects of interatomic interaction, disorder and optical lattice
\cite{Orso}; the second type, by contrast, is unique of disorder
which depletes superfluid fraction even at $T=0$.

Consequently, we follow Ref. \cite{Pitaevskii} and obtain the
analytical expression for the fractional shift in the frequency of
low-lying collective modes
\begin{widetext}
\begin{equation}\label{omega_1}
\frac{\delta \omega }{\omega }=-\frac{\tilde{g}}{2m\omega
^{2}}\left\{ \frac{ \int d\mathbf{r}\tilde{\nabla}^{2} \delta
n^{*}\left( n_{TF}^{2}\frac{\partial k_{int}}{\partial n_{TF}}\delta
n \right) }{\int d\mathbf{r} \delta n^{* }\delta n}+\frac{\int
d\mathbf{r}\left[\tilde{\nabla}^{2}\delta n^{* }\left(
n_{TF}^{2}\frac{\partial k_{dis}}{\partial n_{TF}}\delta n \right)
+n_{n} \tilde{\nabla}\delta n^{* }\cdot \tilde{\nabla}\delta n
\right] }{\int d\mathbf{r}\delta n^{* }\delta n}\right\},
\end{equation}
\end{widetext}
where $\omega$ is the frequency of collective modes obtained from
Eq. ({\ref{mean}}) within the mean-field approximation
\cite{Stringari}.

Eq. (\ref{omega_1}) represents a major result of this paper. In what
follows, we use this equation to derive some analytical results for
phonon propagation and low energy collective modes.

\subsection{Phonon propagation}

An asymptotic analysis of Eq. ({\ref{omega_1}}) allows us to obtain
the sound velocity of the model system in the 3D limit.
Particularly, when the harmonic trap is absent, the model system is
effectively uniform within the context of mass renormalization
theory \cite{Kramer,Hu}. As a result, the $n_{TF}$, $\partial
k_{int}/\partial n_{TF}$, $\partial k_{dis}/\partial n_{TF}$ and
$n_n$ in Eq. ({\ref{omega_1}}) are position-independent and can be
taken out of the integral. Together with Eqs. (\ref{Kint3D}),
({\ref{Kdis3D}}) and ({\ref{Normal3D}}), one obtains the asymptotic
law for the frequency shift in Eq. (\ref{omega_1}) in the 3D regime
\begin{eqnarray}\label{Omega3D}
\frac{\delta \omega}{\omega}&=&-\frac{\tilde{g}(\alpha+\beta+\gamma)
}{4m\omega^2}\frac{n^{3/2}_{TF}\int
d\mathbf{r}\left(\tilde{\nabla}^{2} \delta n^{*}\right)\delta
n}{\int d\mathbf{r}\delta n^{* }\delta n }.
\end{eqnarray}
 Here, $\alpha=(32m^*/3\sqrt{\pi}m)\tilde{a}^{3/2}$,
$\beta=(5m^*/m)\tilde{R}\tilde{a}^{3/2}$ and
$\gamma=-(4\sqrt{\pi}m^*/3m)\tilde{R}\tilde{a}^{3/2}$. Meantime,
solutions of Eq. (\ref{mean}) for this effectively uniform Bose gas
have the form $\delta n\sim e^{i{\bf q}\cdot {\bf r}}$ with a phonon
dispersion $\omega=cq$. Accordingly, we obtain the shift in the
sound velocity
\begin{equation}\label{C3D}
\frac{\delta
c}{c}=\frac{m^*}{m}\left[\frac{8}{\sqrt{\pi}}+\left(\frac{15}{4}-\sqrt{\pi}\right)\tilde{R}\right](n_{TF}\tilde{a}^3)^{\frac{1}{2}}
\end{equation}
with respect to the lattice-modified Bogoliubov value
$c=\sqrt{\tilde{g}n/m^{*}}$. The first term in Eq. ({\ref{C3D}}) is
a consequence of interatomic interaction, which generalizes the
Beliaev result of $\delta c/c=8\sqrt{na^3/\pi}$ \cite{Beliaev} for
the sound velocity to include the effect of optical lattice. On the
other hand, the remaining two terms
$(15/4-\sqrt{\pi})\tilde{R}(n_{TF}\tilde{a}^3)^{1/2}$ are
proportional to $\tilde{R}$, representing the influence of weak
disorder on phonon propagation. As is shown, disorder affects the
sound velocity in two opposite ways: the first is related to quantum
fluctuations that tend to increase sound speed, whereas the second
is related to the depletion of superfluid fraction at $T=0$ which
tends to suppress the phonon propagation. The former effect is also
shared by interatomic interaction, while the latter, by contrast, is
a unique consequence of disorder. In addition, all three terms in
Eq. (\ref{C3D}) demonstrate the same dependence on the gas parameter
$n_{TF}\tilde{a}^3$ as the ground state chemical potential in Eq.
(\ref{E3D}). Eq. (\ref{C3D}) therefore represents generalized LHY
correction to the sound velocity in the presence of optical lattice
and weak disorder.

It's worth mentioning here that one can also employ the
compressibility-based definition for sound velocity
\cite{Kramermass,Liang,Giorgini,Taylor,Menotti,Martikainen} to
investigate the beyond-mean-field-effect induced shift. In our case,
the normal component of fluid is pinned by the quenched disorder and
doesn't directly participate in the propagation of the density
disturbance. The corresponding definition for sound velocity typical
of a superfluid in an optical lattice is therefore
$m^{*}c^2=n_s\partial \mu/\partial n$ \cite{Giorgini}. Here, the
effect of disorder only enters indirectly through the
compressibility $\chi^{-1}=\partial \mu/\partial n$ and the
expression for $n_s$. To carry out the calculation beyond mean-field
approximation, we consider the 3D asymptotic regime where the
chemical potential approaches Eq. ({\ref{E3D}}). The resulting sound
velocity $c_{s}=c+\delta c$ is
\begin{equation}\label{C3D1}
c_{s}=\sqrt{\frac{\tilde{g}n}{m^{*}}}\left\{1+\left[\frac{8}{\sqrt{\pi}}+\left(\frac{15}{4}-\frac{\sqrt{\pi}}{3}\right)\tilde{R}\right]\frac{m^*}{m}(n_{TF}\tilde{a}^3)^{\frac{1}{2}}\right\}.
\end{equation}
The corresponding fractional shift is $\delta
c/c=[8/\sqrt{\pi}+(15/4-\sqrt{\pi}/3)\tilde{R}](m^*/m)(n_{TF}\tilde{a}^3)^{1/2}$.
This result, however, is only consistent with Eq. ({\ref{C3D}}) for
varnishing disorder ($\tilde{R}=0$). In this case, both results
recover corresponding ones in Refs. \cite{Pitaevskii} in free space
and \cite{Orso} in the presence of optical lattice. Nonetheless, in
the presence of disorder, the two differ by an amount
$(2\sqrt{\pi}m^*/3m)\tilde{R}(n_{TF}\tilde{a}^3)^{1/2}$. This
inconsistency in the presence of disorder may be attributed to the
pathology of perturbation theory and its failure (at finite order in
an expansion in powers of $n_{TF}\tilde{a}^3$) to deal with the
total density, condensate depletion and normal density in a
completely consistent way.

\subsection{Collective modes}

Eq. (\ref{omega_1}) allows some direct analysis of respective roles
played by interatomic interaction and disorder on the elementary
excitations of a trapped BEC. The effects beyond mean field on
low-lying collective modes have already been analyzed in free space
\cite{Stringari}, in the presence of optical lattice \cite{Kramer},
and in the presence of disorder (but lattice free) \cite{Falco},
respectively. The following analysis presents a modest
generalization of these previous results to the combined presence of
optical lattice and weak disorder.

According to Eq. (\ref{omega_1}), interatomic interactions will not
affect the frequencies of the so-called surface modes that satisfy
$\widetilde{\nabla}^{2}\delta n^{* }=0$, despite the presence of
optical lattice and disorder. In particular, the dipole mode
corresponds to the center-of-mass motion of the superfluid and will
not be influenced by the interaction, in accordance with the
generalized Kohn's theorem \cite{Pitbook}. Whereas, disorder will
have an immediate effect on surface modes, shifting corresponding
mean-field frequencies in Ref. \cite{Kramer} by an amount
\begin{equation}
\frac{\delta \omega }{\omega }=-\frac{\tilde{g}}{2m\omega
^{2}}\frac{\int d \mathbf{r}n_{n}\tilde{\nabla}\delta n^{\ast }\cdot
\tilde{\nabla}\delta n}{ \int d\mathbf{r}\delta n^{\ast }\delta n}.
\label{Wds}
\end{equation}
In particular, the frequency shift in the dipole mode caused by
disorder, as has been pointed out by Ref. \cite{Falco}, presents a
departure from Kohn's theorem. Here, the result of Eq. (\ref{Wds})
for dipole mode exhibits a magnified departure via the introduction
of optical lattice. Such departure immediately distinguishes the
beyond-mean-field effect due to disorder from that due to
interatomic interaction. By contrast, the compressional modes will
be affected by both disorder and interatomic interaction.

To explicitly observe the effects of disorder, interatomic
interaction and optical lattice on low-lying collective modes, we
consider a deformed harmonic trap in the form
\begin{equation}
V_{ho}(\mathbf{r})=\frac{1}{2}m\omega _{\perp }^{2}r_{\perp }^{2}+\frac{1}{2}
m\omega _{z}^{2}z^{2},  \label{vho}
\end{equation}
where $r_{\perp }=(x^{2}+y^{2})^{1/2}$ is the radial coordinate and
$\omega_\perp$ is the radial trap frequency. The anisotropy is
characterized by the parameter $\lambda =\omega _{z}/\omega _{\perp
}$.

For such anisotropic trap, the zeroth order hydrodynamic Eq.
(\ref{mean}) admits solutions $ \delta {n}(\mathbf{r})\sim
\tilde{r}^{l}Y_{lm}(\theta ,\phi )$ for $m=\pm l$ and $m=\pm (l-1)$
\cite {Stringari} that describes surface excitations, with $
\tilde{r}=\sqrt{m^{\ast }/m}(x^{2}+y^{2})^{1/2}$ being the
lattice-renormalized radial coordinate. The corresponding dispersion
laws are given by $\omega(m=\pm l) =\sqrt{lm/m^{\ast }}\omega
_{\perp }$ and $\omega(m=\pm(l-1))=\left[(l-1)m/m^{\ast
}\omega_\perp^2+\omega_z^2\right]^{1/2}$. The periodic potential
therefore effectively modifies the radial trap frequency
$\omega_\perp$ by a factor $\sqrt{m/m^{*}}$, whereas leaving
$\omega_z$ unaffected. The dipole modes and quadrupole modes are of
particular experimental relevance. The lowest dipole mode
($l=1,m=0$) describes the oscillation of the center of mass along
the z-direction $\delta {n}(\mathbf{r})\sim z$, with the zeroth
order dispersion given by $\omega =\omega _{z}$ which is unaffected
by the lattice. From Eq. ({\ref{Wds}}), the result for the disorder
induced frequency shift in the dipole mode is
\begin{equation}\label{dw10}
\frac{\delta \omega }{\omega }=-\frac{15}{16}\frac{\tilde{R}}{\hbar
d^{2}}\sqrt{\frac{2m\tilde{g}}{n(0)}}\int_{0}^{1}\rho
^{2}\sqrt{1-\rho ^{2}} I\left( \frac{y}{1-\rho ^{2}}\right) d\rho,
\end{equation}
where the variable $y=2t/\tilde{g}n(0)$ with $n(0)$ given by Eq.
({\ref{N0}}) and $I(x)$ is a function of variable $x=y/(1-\rho
^{2})$ given by Eq. ({\ref{I}}). Eq. (\ref{dw10}) shows that the
frequency shift is directly proportional to the strength of disorder
characterized by $\tilde{R}$. For varnishing disorder
($\tilde{R}=0$), there is no frequency shift in the dipole modes, in
conformity with Kohn's theorem \cite{Pitbook}. For the quadrupole
mode $\delta {n}(\widetilde{\mathbf{r}})\sim \tilde{r} ^{2}\sin
^{2}\theta e^{2i\phi }$ with $l=2,m=2$, the mean-field dispersion is
given by $ \omega =\sqrt{2m/m^{\ast }}\omega _{\perp }$. The
corresponding frequency shift is found to be
\begin{equation}\label{dw22}
\frac{\delta \omega }{\omega }=-\frac{21}{8}\frac{\tilde{R}}{\hbar
d^{2}}\sqrt{\frac{2m\tilde{g}}{n(0)}}\int_{0}^{1}\rho
^{4}\sqrt{1-\rho ^{2}} I\left( \frac{y}{1-\rho ^{2}}\right) d\rho,
\end{equation}
which also varnishes in the absence of disorder, consistent with our
previous analysis.

Both Eqs. (\ref{dw10}) and (\ref{dw22}) exhibit a dependence on the
parameter $y=2t/\tilde{g}n(0)$ that characterizes the interplay
between the kinetic energy in a lattice and the mean-field atom-atom
interaction of a trapped gas. In the asymptotic 3D regime where
$y\gg 1$, the function $I(x)$ with $x=y/(1-\rho^2)$ in Eq.
({\ref{dw22}}) decays as $ I(x)\simeq \sqrt{2}/6\pi x$. Accordingly,
one finds the frequency shift in the dipole mode as $\delta \omega
/\omega =-(5\pi ^{3/2}m^{\ast }/64m)\tilde{R}\sqrt{
\tilde{a}^{3}n(0)}$ and in the quadrupole mode as $\delta \omega
/\omega =-(21\pi ^{3/2}m^{\ast
}/256m)\tilde{R}\sqrt{\tilde{a}^{3}n(0)}$. Both results exhibit
dependence on the gas parameter and disorder strength in a similar
way as the normal density does in Eq. (\ref{Normal3D}), which
indicates that the peculiar role of disorder on the surface modes is
closely related to the presence of normal component due to disorder.

To observe the effects beyond mean field on the compressional mode,
we further assume an effective disc-shaped trap $\omega _{z}\gg
\sqrt{m/m^{\ast }} \omega _{\perp }$ for simplicity. In this case,
the zeroth order dispersion for the lowest compressional mode
($n=1,l=0$) is given by $\omega =\sqrt{3}\omega _{z}$ ,
corresponding to a density fluctuation $ \delta n(\mathbf{r})\sim
z^{2}-2\mu /3m\omega _{z}^{2}$. Note that this compressional mode is
only determined by the trap frequency $\omega_z$, thereby unaffected
by the lattice applied along the $x-y$ plane. It follows from Eq.
({\ref{omega_1}}) that the frequency shift in the mode ($n=1,l=0$)
is
\begin{equation}\label{Dwn1}
\frac{\delta \omega }{\omega }=\frac{21}{512}\frac{1}{\hbar
d^{2}}\sqrt{ \frac{2m\tilde{g}}{n(0)}}\left[ K(y)+\tilde{R}[(Z( y)
-T(y)] \right].
\end{equation}
Here, $y=2t/\tilde{g}n(0)$, and $K(y)=(16/\pi
)y^{1/2}\int_{0}^{1}\rho ^{2}(\rho ^{2}-1)D(x) d\rho$ with
$x=y/(1-\rho^2)$ and $D(x)=x^{3/2}f^{\prime \prime
}(x)-3f(x)/4\sqrt{x}$. Using the same definitions for variables $x$
and $y$, the functions $Z(y)$ and $T(y)$ are defined as
$Z(y)=(8\sqrt{2}/\pi) y^{1/2}\int_{0}^{1}\rho ^{2}(\rho ^{2}-1)G(x)
d\rho$ where $G(x)=y^{3/2}Q^{\prime \prime }(x)-3Q(x)/4\sqrt{x}$,
and $T(y)=32\int_{0}^{1}\rho ^{4}\sqrt{1-\rho ^{2}}I(x) d\rho$ where
$I(x)$ follows Eq. (\ref{I}). Eq. ({\ref{Dwn1}}) confirms our
previous statement that the compressional mode is affected by both
repulsive atom-atom interaction and disorder. Specifically, the
first term in Eq. ({\ref{Dwn1}}) represents the contribution from
interatomic interaction, whereas the latter two are consequences of
disorder. Moreover, Eq. ({\ref{Dwn1}}) shows that disorder has two
opposite effects on compressional excitations: the induced quantum
fluctuation tends to increase excitation frequencies, whereas the
depletion of superfluid fraction tends to decrease the excitation
energy. This latter effect exhibited by the term proportional to
$T(y)$, in particular, is unique of the effect of disorder on
compressional modes. For vanishing disorder $\tilde{R}=0$, Eq.
({\ref{Dwn1}}) recovers the corresponding result in Ref.
\cite{Orso}.

The parameter $y=2t/\tilde{g}n(0)$ controls the dimensional
crossover of the frequency shift in Eq. ({\ref{Dwn1}}). In the asymptotic 3D regime corresponding to $y\gg 1$, functions $f(x)$, $Q(x)$ and $I(x)$ with $x=y/(1-\rho^2)$ respectively decay as $f(x)\simeq 1.43/\sqrt{x}-16\sqrt{2}/15\pi x$, $Q(x)\simeq -1/%
\sqrt{\pi }x$ and $I(x)=\sqrt{2}/6\pi x$. Hence Eq. ({\ref{Dwn1}})
asymptotically approaches
$\delta\omega/\omega=35\sqrt{\pi}m^{*}/128m\left[1-3/5\tilde{
R}\left(25\sqrt{\pi}/64-\pi/8\right)\right]\sqrt{\tilde{a}^{3}n(0)}$.
This result provides a further generalization by disorder of the
frequency shift caused by the lattice-generalized LHY correction in
Ref. \cite{Orso}. Moreover, this result shows that the
disorder-induced frequency shift is of the same order of magnitude,
but with an opposite sign, compared to that due to atom-atom
interaction.

\section{Possible experimental scenarios and conclusion}

Central to test the validity of the hydrodynamic picture proposed in
this paper are the experimental abilities to measure the superfluid
density and condensate density respectively. For a gaseous BEC, the
condensate fraction is readily measured through mapping of
occupation numbers in the momentum space to real space by expansion
imaging \cite{Dalfovo}. Whereas, to determine the superfluid
fraction, one can use a method proposed in Ref. \cite{Cooper}.
There, a BEC is beard to an optically-induced vector potential,
which simulates a uniform rotation of BEC. As a result, the normal
fluid picks up non-zero angular momentum, whereas the superfluid
acquires no angular momentum. Thus the spectroscopy can be used to
measure the net change in the angular momentum of the fluid, and
therefore the superfluid fraction.

Another difficulty that may arise is how to experimentally measure
the beyond-mean-field correction to the equation of state due to the
interatomic interaction and disorder. The concern is as follows
\cite{Xu}: for a harmonic trap, the quantum depletion cannot be
observed during the ballistic expansion in the typical
Thomas-regime. Because the mean-field energy is much larger than the
trap frequency, the cloud remains locally adiabatic during the
expansion. The condensate at high density transforms adiabatically
into a condensate at low density with diminishing quantum depletion.
The above concern is true for a harmonically trapped BEC. However,
it can be ruled out by introducing an optical lattice. In general,
the confinement frequency at each lattice site far exceeds the
interaction energy, and the time-of-flight images are essentially a
snapshot of the momentum distribution at the time of the lattice
switch-off, thus allowing for a direct observation of the quantum
depletion.

Upon overcoming the above two difficulties, the experimental
realization of our scenario amounts to controlling three parameters
whose interplay underlies the physics of this work: the strength of
an optical lattice $s$, the mean-field interaction between bosonic
atoms $\tilde{g}n_{TF}$, and the strength of disorder $\tilde{R}$.
All these quantities are experimentally controllable using
state-of-the-art technologies. The interatomic interaction can be
controlled in a very versatile manner via the technology of Feshbach
resonances \cite{Fesh}. In the typical experiments to date, the values of ratio $\tilde{g%
}n_{TF}/E_R$ range from $0.02$ to $1$ \cite{Morschrev,Blochrev}. The
depth of an optical lattice $s$ can be changed from $0E_R$ to
$32E_R$ almost at will \cite{Greiner}. The above scenario for
disorder can be realized in cold atomic systems using several
methods in a controlled way. They include applying optical
potentials created by laser speckles or multi-chromatic lattices
\cite{Damski,Roth,Lye,White}, introducing impurity atoms in the
sample \cite{Ospelkaus} and manipulating the collision between atoms
\cite{Gavish}. Therefore, the phenomena discussed in this paper
should be observable within the current experimental capability.
We emphasize here that the work presented in this
paper is restricted to weak disorder and weak interatomic interaction.
For further investigations in the presence of stronger inter-atomic interaction
or disorder, the path-integral Monte Carlo simulation is a reliable method \cite{Pilati}.

In summary, we explore the combined effects of weak disorder and a
2D optical lattice on the collective excitations of a harmonically
trapped BEC. Accordingly, we extend the hydrodynamic equations for
superfluid \cite{Kramer} for a 2D optical lattice to include the
effects of weak disorder. For small amplitude excitations, the
corresponding equations are characterized by the disorder-depleted
superfluid fluid density, a renormalized interatomic interaction and
an effective mass capturing the effect of periodic potential. Using
these hydrodynamic equations, we calculate the collective
frequencies beyond the mean-field approximation. Our results reveal
peculiar role of disorder, interplaying with optical lattice and
atom-atom interactions, on the elementary excitations of a trapped
BEC. In particular, we present detailed analysis for consequences of
disorder on phonon propagation and lattice-modified surface modes.
Conditions for possible experimental realization of our scenario are
also proposed.\\

\textbf{Acknowledgements} We thank Biao Wu for helpful discussions.
YH and BBH are supported by the Hongkong Research Council (RGC) and
the Hong Kong Baptist University Faculty Research Grant (FRG). ZXL
is supported by the IMR SYNL-TS Ke Research Grant.

\end{document}